\documentclass[journal]{IEEEtran}

\usepackage{subfig}
\usepackage{amsmath,amssymb,epsfig,bbm,url}
\usepackage{graphicx,color}
\usepackage{cite}
\title{A Phase Lookahead DTC for Fast Settling Switched Loop DPLL}

\begin{document}
\author{Pallavi~Paliwal,
        Vivek~Yadav,
        and~Shalabh~Gupta% <-this % stops a space
\thanks{The authors are with the Department
of Electrical Engineering, Indian Institute of Technology Bombay, Mumbai 400076, India (e-mail: pallavi.paliwal@iitb.ac.in; shalabh@ee.iitb.ac.in).}}% <-this % stops a space

% \markboth{IEEE JOURNAL OF SOLID-STATE CIRCUITS}{Paliwal \MakeLowercase{\textit{et al.}}: Dual-phase DDS based DTC with variable phase-lookahead}

\maketitle
% The command maketitle displays the title and author information from preamble part here.

%

\begin{abstract}
 
In most digital-to-time converter (DTC) based applications, apart from maintaining low integral non-linearity (INL), it is also required of the system to achieve a wide frequency translation range. To achieve this performance, we present a dual-phase  direct digital synthesizer (DDS) based DTC with phase-lookahead mechanism. The proposed technique of variable phase-advancement enhances the frequency translation range, without excessive power consumption. A 5-GHz digital phase locked loop (DPLL) with switched loop, incorporating this DDS based DTC, is implemented in CMOS65\,nm-LL technology. The proposed DDS based DTC is able to perform fractional shift upto $\pm$80\,MHz with 100\,MHz reference clock, using 3\,mW of power from 1.2\,V supply. A simple look-up table based foreground-calibration of phase-to-amplitude converter (PAC) in DDS improves the peak INL of the DTC to 0.25\,ps. Hence, with the proposed DTC and a proportional-integral-derivative (PID) controller based loop, we are able to achieve a low-jitter fractional-N DPLL with fastest  settling time of 1-$\mathrm{\mu} s$ reported until now for fractional-N PLLs.
\end{abstract}


%********************************************************************
% Introduction
%*********************************************************************************8

\section{Introduction}

\begin{figure}[h]
\centering
{
\subfloat[]{
\includegraphics[scale=0.6]{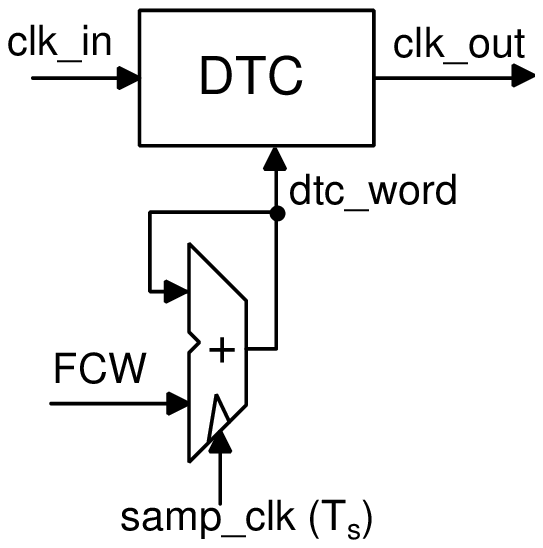}}
\subfloat[]{
\includegraphics[scale=0.6]{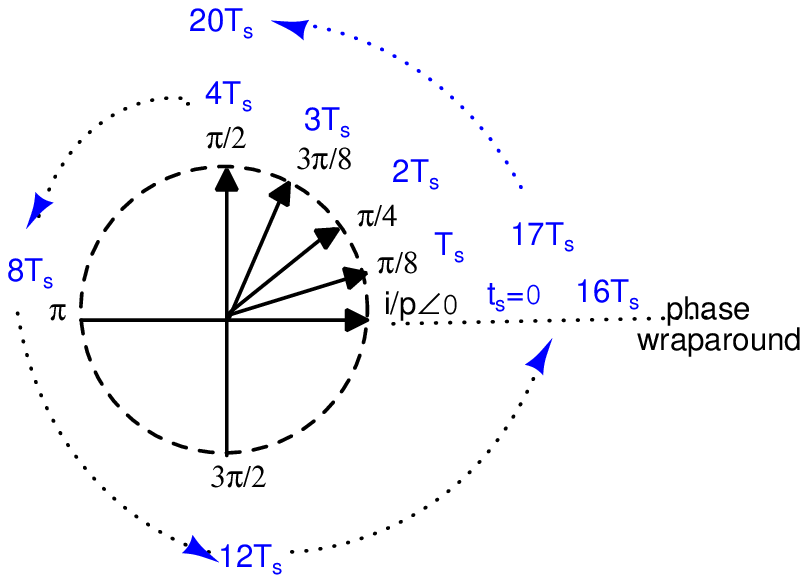}}
}
\caption{(a) DTC with constantly increasing control word (\textit{dtc\_word}) from accumulator, generates infinite delay-range over time. (b) Phasor of 4-bit Accumulator+DTC output having incremental phase-shift/delay corresponding to $FCW=1$. }
\label{fig:phase_wraparound}
\end{figure}

\begin{figure}[h]
\centering
{
\subfloat[]{
\includegraphics[scale=0.7]{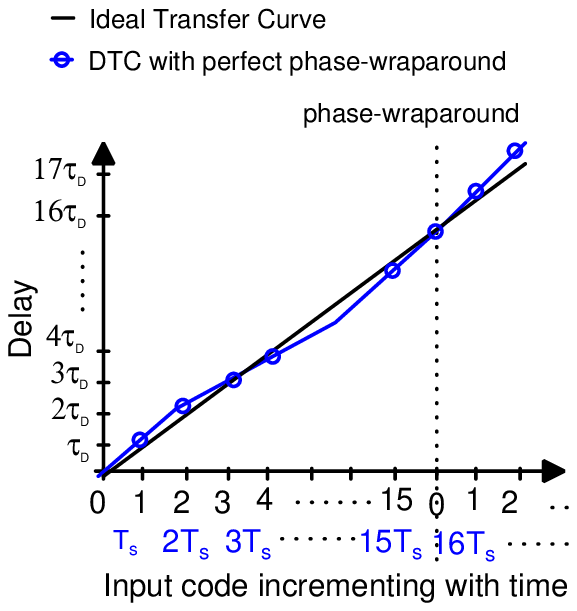}}
\subfloat[]{
\includegraphics[scale=0.7]{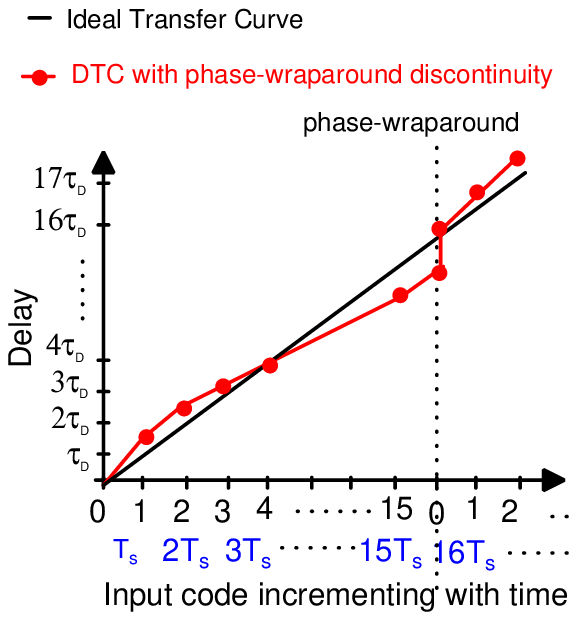}}
}
\caption{Infinite delay-range over time characteristic in a DTC, with system having information of (a) end-points of the transfer-curve; (b) only initial point of the transfer curve.}
\label{fig:dtc_inf_delay}
\end{figure}

In current generation digital phase-locked loop (DPLL) architectures, digital-to-time converters (DTCs) have been extensively utilized for efficient fractional division. The DTC, as a standalone building block, generates the desired delay in a finite delay-range, based on the programmed control word. However, fractional-N PLLs \cite{ref:dpll_dtc_pi,ref:dpll_pi, Ref1_Extended_divider,Ref7_vlsi_conf_dpll} operate with the DTC emulating the property of infinite-delay-range over the time, by generating the desired frequency offset using continuously incrementing output from a phase accumulator.    Figure \ref{fig:phase_wraparound} shows a DTC+Accumulator generating a constantly rotating phasor, which leads to the required frequency translation of the DTC input ($clk\_{in}$).  In the DTC+Accumulator implementation, as the time progresses, the accumulator output (\textit{dtc\_word}) increments with the programmed frequency control word   (\textit{FCW}) at the sampling clock rate ($T_s$), thus generating an incremental delay ($\tau_d$) in the DTC input signal. When the accumulator output overflows, the DTC signal also undergoes a phase-wraparound, as shown in Fig. \ref{fig:phase_wraparound}(b). The DPLL  performs better in terms of spur rejection if the DTC possess inherent phase-wraparound feature, which effectively provides a seamless  infinite-delay-range emulation \cite{Ref8_High_resolution_DTC}.   Figure \ref{fig:dtc_inf_delay} highlights that a DTC has perfect phase-wraparound as an inherent feature, if the gain (or) endpoints of the transfer characteristics are well-defined in the system.

\begin{figure}[h]
\centering
\includegraphics[scale=0.6]{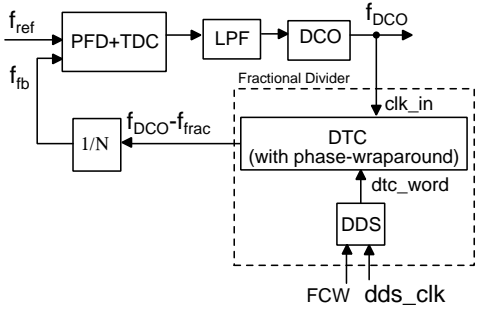}
\caption{DPLL block diagram with DDS based DTC as a ``plug-n-play'' block for fractional division \cite{Ref10_A_low_jitter-DTC}.}
\label{fig:dtc_pll}
\end{figure}

The transfer characteristics in Fig. \ref{fig:dtc_inf_delay} highlights that a DTC implemented with two or more phases of input signal, has to be calibrated only for nonlinearities, since the gain is deterministic with known endpoints of the transfer curve. On the contrary, a DTC receiving single phase of incoming signal as an input, has to be calibrated for both gain and non-linearities in the transfer characteristics. The drawback is that the DTCs requiring frequency-dependent gain calibration have convergence time in the order of tens of microseconds, thus slowing down the system speed. The perfect phase-wraparound feature of a DTC can help in overcoming the
above mentioned problem and achieving instantaneous frequency switching. A direct digital synthesizer (DDS) based DTC, working on this calibration-free infinite delay-range property, acts as a simple plug-n-play fractional divider in a DPLL, as shown in Fig. \ref{fig:dtc_pll}. We have shown a lock-time enhanced DPLL architecture incorporating DDS based DTC in \cite{ref:proposed_dpll}, which uses a switched loop with variable phase-detection and proportional-integral-derivative (PID) controller based finite state machine (FSM) \cite{ref:dpll_fom} to achieve a record low settling time.

For attaining low jitter, most of the DDS based DTCs incorporate harmonic-rejection, polyphase mixing and/or anti-aliasing filtering\cite{ref:dds_pll}, which result in high power dissipation. Even with these filtering techniques, the DTC jitter does not reach sub-picosecond range, and the frequency translation range remains limited to the Nyquist rate.

In this work, we have proposed a dual-phase DDS based DTC architecture with a phase look-ahead mechanism, to achieve a wide frequency-translation range. Instead of power-consuming high-order filters, the proposed architecture uses time and phase synchronized DDS array for undesired spurs rejection. While the usage of a DDS array for uncorrelated noise rejection has been discussed in \cite{ref:dds_array}, this work uses the concept of multi-phase DDS with phase-advanced information for additional attenuation of highly correlated spurs. In addition, by carefully choosing the output sampling edge, we are able to achieve a DTC with very low 
INL, without even using any bandpass or polyphase filtering techniques. 

Towards developing a DDS based DTC system having wide-range frequency modulation with low jitter, we  discuss suitable DTC variants and concept of multi-phase DDS approach in Section \ref{sec:DTC_arch} and Section \ref{sec:DDS_overview}, respectively. The frequency-range enhancement techniques for dual-phase DDS based DTC system have been discussed in Section \ref{sec:dual_phase_dtc}. Section \ref{sec:alias_shift} analyzes the shift in alias-frequency with the dual-DDS array, and Section \ref{sec:dtc_implementation} highlights the reasons for spur-origin in the DTC implementation. Section \ref{ref:switched_loop} sets forth the basis of the switched-loop DPLL system which is capable of achieving fastest-reported  settling time with its novel PID controller. Section \ref{sec:dtc_meas} presents the performance of proposed DDS based DTC as a fractional divider in the aforementioned 5-GHz DPLL, implemented in CMOS65\,nm-LL technology. Section \ref{sec:calibration} shows further improvement in the DTC linearity with a calibrated pre-distortion applied to the DDS look-up table (LUT). The performance comparison of DTC variants in Section \ref{sec:perf_cmp} manifests that the proposed dual-phase DDS based architecture operates with state-of-the-art INL of 0.25\,ps with optimal power consumption, without the need of background calibration.

%**************************************************************************
% Choice of DTC Architectur
%**************************************************************************  
\section{Choice of DTC Architecture}
\label{sec:DTC_arch}
\begin{figure}[!h]
\centering{\includegraphics[scale=0.5]{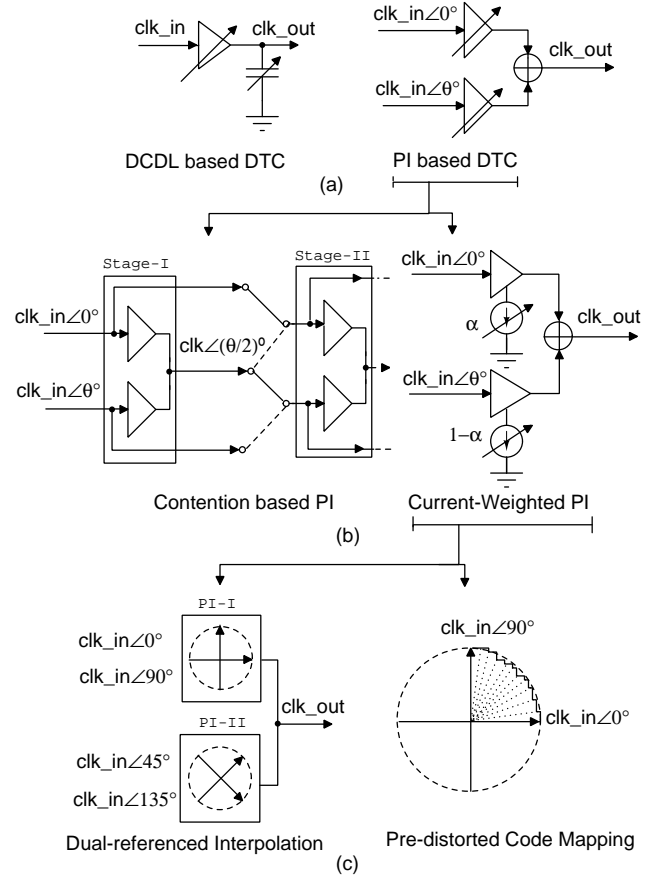}}
\caption{(a) Types of DTCs; (b) Types of Phase Interpolators \cite{ref:dpll_dtc_pi}; (c) Linearization techniques in current-weighted phase interpolators.}
\label{fig:dtc_type}
\end{figure}

DTCs are generally implemented using either a digitally controlled delay line (DCDL) or a phase interpolator (PI), as shown in Fig. \ref{fig:dtc_type}(a). A DCDL, using single input-phase for delay generation, has to be calibrated for both gain and non-linearities due to PVT mismatches and random variations. A phase interpolator, on the other hand, uses two-phases of input signal to generate an intermediate-phase. Therefore, gain in the case of a phase-interpolator is fixed and only non-linearities in the transfer-characteristics needs to be calibrated. Due to a deterministic gain, the perfect phase-wraparound property inherent to a phase interpolator easily fulfills the infinite delay-range requirement on DTC. A DCDL, on the contrary, needs continuous feedback from the system towards background calibration, especially to avoid drastic step-change during wraparound from maximum to minimum DTC control word value. In PLLs involving DCDL based DTCs, the convergence of the DTC-gain calibration loop also affects the settling response of the main loop. This convergence time of the calibration loop deteriorates further, when the DPLL has to lock to a near-interger channel (i.e. small fractional control word) since the correlation loop and the main-loop disturb each other \cite{ref:ulp_ss_dpll}. Though \cite{ref:ulp_ss_dpll} proposes a variable-preconditioned least-mean square (LMS) algorithm for fast calibration in a DTC, the convergence speed still remains in the order of 40\,$\mathrm{\mu}$s.

Thus, towards a fast-settling DPLL design, usage of phase-interpolator based DTC as a fractional divider turns out to be an easier option. Most popular variants of phase-interpolators are (i) Contention-based PI and (ii) Current-weighted PI, as shown in Fig. \ref{fig:dtc_type}(b). A contention-based PI involves multiple inverters sharing a common output which leads to an additional short-circuit error degrading the INL \cite{ref:dpll_dtc_pi}. Inspite of a short-circuit error suppression technique shown in \cite{ref:dpll_dtc_pi}, INL of the DTC still remained limited to 1.4\,ps. Another tradeoff of concern is that the edge-rate degraded in contention-PI for better linearity (to avoid effect of time-varying nonlinear resistance) results in a lower noise immunity \cite{ref:dtc_hf}. The pipelined-PI in \cite{ref:dpll_dtc_pi}, for instance, highlights that if additional interpolator stages are added for finer resolution (i) intrinsic delay increases and (ii) INL degrades, since any phase imbalance is propagated and could be amplified in subsequent stages. In other variants of PI also, for instance in \cite{ref:apf_pi} employing polyphase-filtering based PI, the trade-off between power and linearity is visible.

 This work explores a fractional-divider architecture based on a current-mode PI, with the aim of achieving low-jitter, low-power and instantaneous fractional frequency generation in the DPLL feeedback path. Equation (\ref{eqn:in_band_spur}) \cite{ref:in_band_spur} shows that the non-linearities in the employed DTC directly reflects as fractional spurs at the DPLL output. Therefore, improving INL of the proposed DTC is of paramount important. 

\begin{equation}
\label{eqn:in_band_spur}
 L=\frac{\pi^2}{4}\left(\frac{INL_{pp}}{T_{CKV}}\right)^2
\end{equation}
 where, $L$ = In-band spur level.\\

The linearization techniques of (i) nonlinear code mapping in PI, (for instance, octagonal-rotator in \cite{ref:octagonal_rotator}) and (ii) dual-referenced PI \cite{ref:dtc_hf}, shown in Fig. \ref{fig:dtc_type}(c), are widely used to improve the DTC INL performance. In the implemented design, LUT of the dual-phase DDS is pre-distorted to equalize the nonlinearities in the DTC. With this linearization technique, the proposed DTC based fractional divider is able to achieve a low rms-jitter of 0.19\,ps, without impacting the settling response of the employing DPLL loop.

%**************************************************************************
%DDS BASED DTC SYSTEM OVERVIEW
%**************************************************************************
\section{DDS Based DTC System Overview}
\label{sec:DDS_overview}
In an offset DPLL system, the DDS generates the fractional frequency  ($\omega_{{frac}}$) using which the DTC, acting as a simple phase-rotator, shifts the frequency ($\omega_{LO}$) of incoming oscillator signal. This section highlights the characteristics of single-phase DDS and DDS-array based  DTC which could be used as a standalone phase rotator, in contrary to DCDL based DTCs needing calibration feedback from the external system.
\subsection{Single-phase DDS based DTC}

\begin{figure}[h]
	\centering
\includegraphics[scale=0.5]{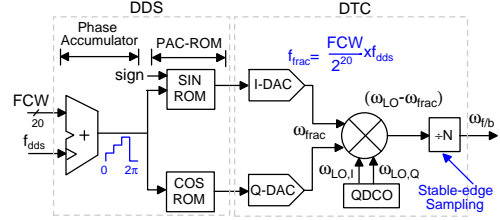}
	\caption{Block diagram of a conventional DDS based DTC system, employed as a fractional divider in the DPLL feedback path.}
	\label{fig:Basic_DDS_DTC}
	\end{figure}

A conventional DDS based DTC architecture in Fig. \ref{fig:Basic_DDS_DTC} consist of (i) a phase accumulator, (ii) a phase to amplitude convertor (PAC) implemented with a read only memory (ROM) followed by (iii) a digital-to-analog converter (DAC) and (iv) a mixer, as a fractional frequency divider in the PLL. In Fig. \ref{fig:Basic_DDS_DTC}, based on the programmed frequency controlled word (FCW) and sampling clock ($f_{ref}$),  the DDS based DTC system modulates the incoming  quadrature digitally controlled oscillator signal (QDCO), for removal of fractional frequency ($f_{frac}$) component from the PLL feedback path. 
 
A single-phase DDS based DTC architecture has restricted performance in terms of (i) limited fractional frequency range due to aliased component, and  (ii) increased output jitter due to harmonics generated with large quantization step. 
%--------------------------------------------------
% Multi-phase DDS based DTC
%--------------------------------------------------

\subsection{Multi-phase DDS based DTC}

\begin{figure}[h]
	\centering
{
 	\includegraphics[scale=0.58]{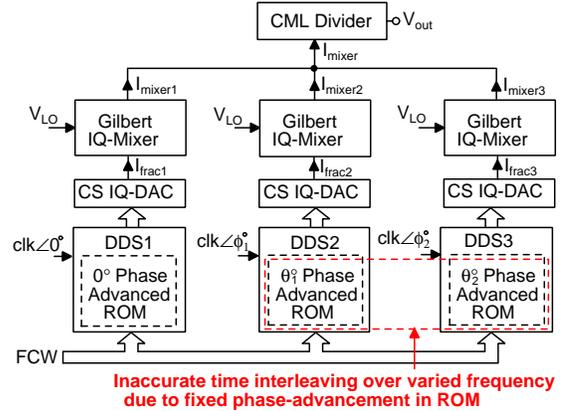}
}
	\caption{ Multi-phase DDS based DTC with phase-advanced ROMs and incrementally delayed sampling clocks.}
	\label{fig:Multi_DDS}
	\end{figure}

To mitigate the issue of harmonics and aliased component restricting the fractional frequency range, we proposed a multi-phase DDS based DTC architecture in \cite{Ref10_A_low_jitter-DTC}  as an improvement over a conventional DDS  architecture. The architecture in Fig. \ref{fig:Multi_DDS} uses multiple DDSs with its ROM being phase-advanced  and sampling-clock being delayed to generate an interpolated waveform analogous to a second-order hold response. The summed-output from multi-DDSs arrangement reduces harmonics and aliased components by avoiding steep transition using phase-advanced information in the system. 
	
Though the phase-shifted DDS array is able to extend the frequency modulation range beyond the Nyquist rate of sampling clock, the layout complexity increases while trying to avoid the noise generated from cross-talk and delay imbalance at multiple DDSs output. To circumvent the need of generating a matched layout for a multi-phase DDS based DTC, this work derives frequency-modulation range enhancement technique for a dual-phase DDS based DTC system. A brute-force approach of using a fixed phase-shift in the ROM and clock of a DTC system, analogous to Fig. \ref{fig:Multi_DDS} with DDS3 cell removed, still has limitation in terms of achievable fractional frequency range. The concept behind the limitation is that a fixed phase advancement in the DDS-ROM would not lead to the interpolation at correct intermediate-phase for all possible values of output frequency range.

%******************************************************************************
% Performance enhancement of dual DDS based DTC architecture
%*****************************************************************************
\section{Performance enhancement of Dual-phase DDS based DTC architecture}
\label{sec:dual_phase_dtc}
A dual-phase DDS based DTC could cover a maximum frequency modulation range, if the phase look-ahead based waveform interpolation  is applied at the correct time instant in the DDS-generated waveform. However, for different ranges of DDS generated frequencies, the ROM requires different amount of phase-advancement   for correct look-ahead interpolation. Thus, instead of a fixed phase-advancement, the dual-phase DDS based DTC needs a variable phase-shift in the additional DDS, with the phase-shift value depending on the programmed FCW. 

%	\begin{figure}[h]
%	\centering
% 	\includegraphics[scale=0.4]{././dtc_wrkg}
%	\caption{b.clock Reference input to DDSs.}
%	\label{fig:dtc_wrkg}
%	\end{figure}

% \subsection{Variable phase-advanced dual-phase DDS based DTC}

\begin{figure}[h]
% 	\centering 
{
 	\includegraphics[scale=0.245]{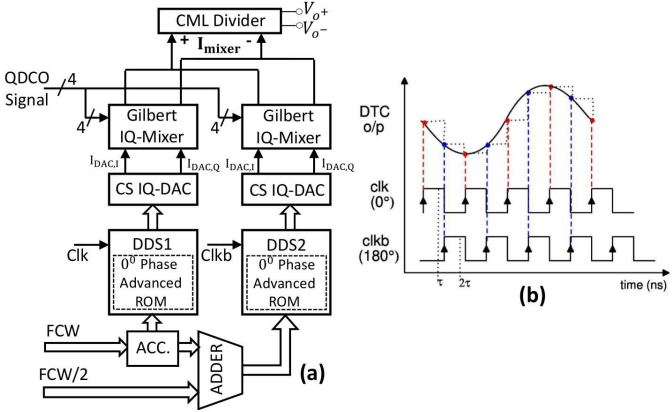}
%  \subfloat[]{\includegraphics[scale=0.5]{./dualphase_dds_add_v2.eps}}
% \subfloat[]{\includegraphics[scale=0.36]{./dtc_wrkg.eps}}
}	\caption{(a) Variable phase-advanced dual-phase DDS based DTC with fixed time interleaving at intermediate-phase of 180$^\circ$. (b) Complementary-clock edges sampling the two DDSs allow mid-point interpolation in the combined DDS-waveform.}
	\label{fig:adderDDS_DTC}
	\end{figure}

With the aim of introducing correct interpolated points in the DDS generated waveform for the required frequency range, this work proposes advancing the FCW input to DDS2-ROM shown in Fig. \ref{fig:adderDDS_DTC}(a), rather than hard-coding fixed phase-advanced values in the ROM.  In the proposed modifications to dual-phase DDS based DTC, while the input address word for DDS2-ROM is advanced by FCW/2, the DDS2-clock (\textit{clkb}) is also phase-shifted by 180$^\circ$. With the suggested operating principle, the output of both the DDSs are exactly time and phase interleaved, as shown in Fig. \ref{fig:adderDDS_DTC}(b). The in-phase relation between the two DDSs cause the output of corresponding mixers to be additive in nature.

\begin{figure}[htb]
	\centering 
	\includegraphics[scale=0.42]{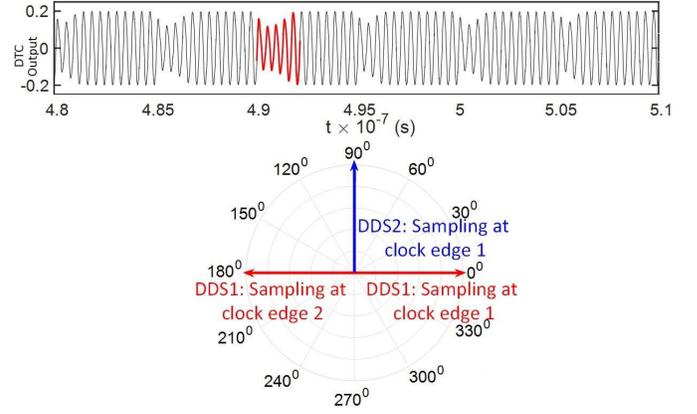}
	\caption{DTC output swing reduction with inversion of phase-relation in-between the output of dual-DDSs.}
	\label{fig:DDS_DTC_Adder_limitation}
	\end{figure}

 The drawback still pertaining to the architecture in Fig. \ref{fig:adderDDS_DTC} is that the outputs of both the DDSs are switching alternatively, causing an instantaneous phase change in-between them. Figure  \ref{fig:DDS_DTC_Adder_limitation} shows that at each sampling instant of DDS-clock (\textit{clk}), output of one DDS switches its phase from lagging to leading with respect to the other. This phenomena results in an increased output jitter ($\approx$3.4\,ps) for fractional frequency range ($>$ 45\,MHz with $f_{clk}$=100\,MHz) near the Nyquist rate. 

%-----------------------------------------------------------------
%Proposed Dual-Phase DDS based DTC Architecture
%----------------------------------------------------------------
% \subsection{Proposed dual-phase DDS based DTC architecture} 
% \label{sssec:proposed_arch}

\begin{figure}[htb]
	\centering
 \includegraphics[scale=0.58]{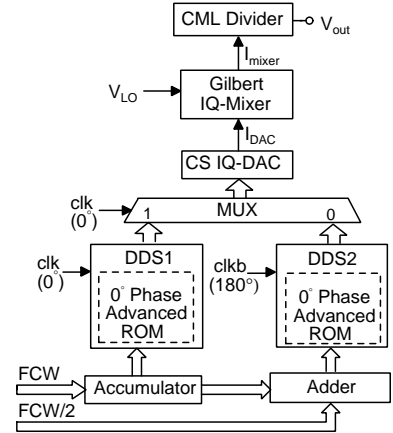}
	\caption{Proposed dual-phase DDS based DTC system with variable phase-advancement and fixed time-interleaving using multiplexer.}
	\label{fig:Dual_DDS_DTC_MUX}
	\end{figure}

As a low-jitter technique for frequency modulation range enhancement, this work proposes a variable phase-advanced dual-phase DDS based DTC architecture in Fig. \ref{fig:Dual_DDS_DTC_MUX}, with a multiplexer employed for fixed time-interleaving between the two DDSs. The multiplexer, with DDS clock ($clk$) as the select signal, combines the output of two DDSs with exact time interleaving of  $T_s$/2 (where $T_s$ is the DDS clock period). Thus, the time-interleaving of phase-shifted DDS output with a multiplexer avoids the instability inherent to the current-mode summation of DDSs output.

%-----------------------------------------------------------------
% Alias-frequency shift with the proposed DTC
%----------------------------------------------------------------

\section{Alias-frequency shift with the proposed DTC}
\label{sec:alias_shift}
The dual-phase DDS based DTC in Fig. \ref{fig:Dual_DDS_DTC_MUX} emulates a single-phase DDS based DTC operating at double the sampling rate ($2f_{ref}$). Therefore, the aliased component corresponding to $f_{ref}$ frequency at  $f_{ref} + f_{frac}$ and $f_{ref} - f_{frac}$ locations are intuitively expected to be absent in the DTC output. This cancellation of nearby aliased component is proved as follows, in similar lines as \cite{Ref11_time_interleaving}.

Let an ideal signal required to be generated from a DDS be represented as $x(t)$, and the actual sampled output signal of DDS be denoted by $x_i[n]$, where, '$i$' refers to the path index in a DDS array and $T_s$ is the DDS clock period. The sampled output generated by DDS1 and DDS2 can be written as,

 \begin{equation}
\label{eqn:sampled_op}
\begin{aligned}
{x_{1}}[n] &= {x(t)}{\delta(t - nT_s)}, \\
{x_{2}}[n] &= {x(t)}{\delta(t - nT_s - T_s/2)}.
\end{aligned} 
\end{equation}

%=  {x(\frac{2t - \tau}{2})}{\delta(t - n\tau)}
\noindent The Fourier transform of (\ref{eqn:sampled_op}) is given by

\begin{equation}
\label{eqn:fourier}
\begin{aligned}
{X_{1}}[e^{ j\omega}] &=\frac{1}{T_s} \sum_{k = - \infty}^{k = + \infty} X\left(j\left(\omega - \frac{2\pi k}{T_s}\right)\right), \\
{X_{2}}[e^{ j\omega}] &=\frac{1}{T_S} \sum_{k = - \infty}^{k = + \infty} X\left(j\left(\omega - \frac{2\pi k}{T_s}\right)\right) e^{j\frac{\omega T_s}{2}}e^{-j\frac{2\pi  k}{N}}.
\end{aligned}
\end{equation}

\noindent Equation (\ref{eqn:fourier}) is incomplete because DDS2 not only operates on a phase-shifted clock, but also has the PAC-ROM phase-advanced by FCW/2 value. This phase-advancement translates into $T_s/2$ advancement in time-domain, thus, a modified fourier transform for DDS2 can be represented as 

\begin{equation}
\label{eqn:dds2}
{X_{2}}[e^{ j\omega}] =\frac{1}{T_s} \sum_{k = - \infty}^{k = + \infty} X\left(j\left(\omega - \frac{2\pi k}{T_s}\right)\right)e^{j\frac{\omega T_s}{2}}e^{j\frac{\omega T_s}{2}} e^{-j\frac{2\pi k}{2}}.
\end{equation}

\noindent Summing the fourier transform of dual-phase DDS output gives

\begin{equation}
\label{eqn:dds3}
{X}[e^{ j\omega}] =\frac{1}{T_s} \sum_{k = - \infty}^{k = + \infty} X\left(j\left(\omega - \frac{2\pi k}{T_s}\right)\right) \left(1 +  e^{-j\frac{2\pi k}{2}}\right)
\end{equation}	

where,

\begin{equation*}
 \left(1 +  e^{-j\frac{2\pi k}{2}}\right) =
 \begin{cases}
0 \quad  \forall \quad\:  k  \: \neq \: \left(0, 2, 4, \dots\right)\\
2 \quad  \forall \quad\:  k  \: = \: \left(0, 2, 4, \dots\right)\\
\end{cases}
\end{equation*}	

\begin{figure}[!h]
		\centering 
{
\subfloat[]{\includegraphics[scale=0.6]{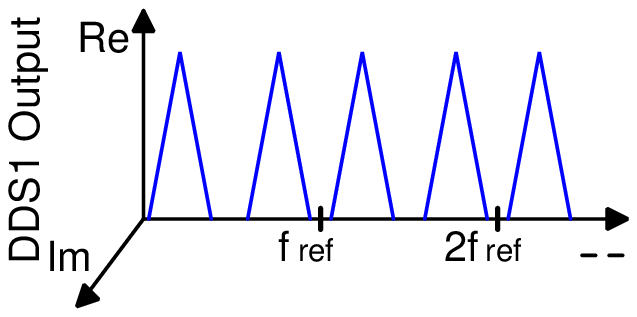}}
\subfloat[]{\includegraphics[scale=0.5]{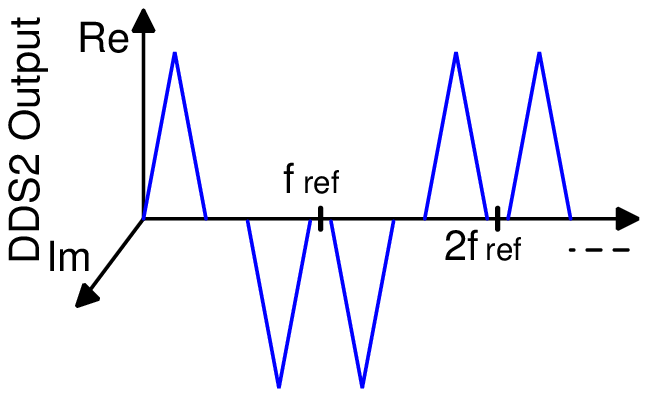}}\\
\subfloat[]{\includegraphics[scale=0.57]{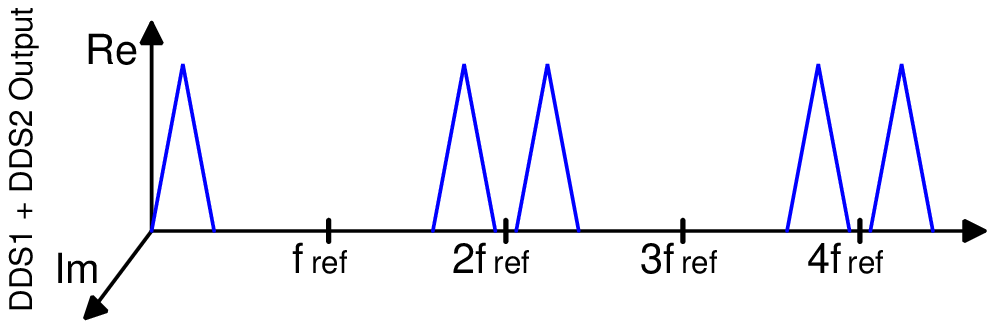}}
}
  		 \caption{Image replica cancellation in the proposed dual-phase DDS based DTC architecture.}
  		 \label{fig:picall}
		\end{figure}

\noindent Equation (\ref{eqn:dds3}) suggests that the odd-ordered image replica components get cancelled with the proposed DTC implementation, which can also be observed from Fig. \ref{fig:picall} and simulated response in Fig. \ref{fig:DFT_Dual_DDS_based_DTC_with_MUX}. Hence with a first-order aliased component rejection, the frequency modulation range of DDS based DTC extends beyond the Nyquist rate. 

%--------------------------------------------------------------
% Reason for Fractional SPur Generation
%--------------------------------------------------------------
\section{Dual-Phase DDS based DTC implementation}
\label{sec:dtc_implementation}
\begin{figure}[!h]
	\centering
 	\includegraphics[scale=0.5]{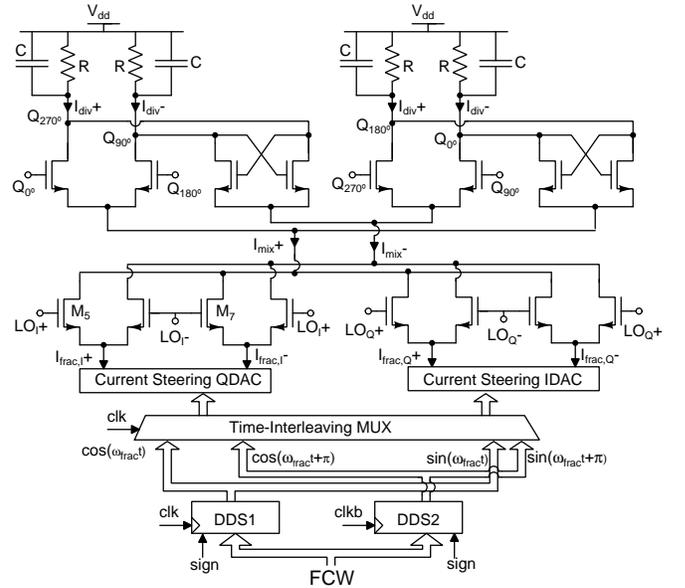}
	\caption{Current-weighted DTC with input from time-interleaved dual-phase DDS.}
	\label{fig:dtc_implementation}
	\end{figure}

The dual-phase DDS based DTC implementation in Fig. \ref{fig:dtc_implementation} uses a  cascade-arrangement with current-mode signaling for DAC, mixer and succeeding current-mode logic (CML) divider in the DPLL feedback path. The current-mode cascaded arrangement prevents jitter increment due to transconductance non-linearities, and reduces power consumption with current reuse mechanism \cite{Ref8_High_resolution_DTC}. The 8-bit current-steering DAC is implemented using segmented architecture with lower 4-bit binary-weighted DAC and upper 4-bit thermometer-weighted DAC. The DDS involves a sign bit to complement the DDS output for generating frequency modulation in both ($f_{LO}+f_{frac}$) and ($f_{LO}-f_{frac}$) range, where LO is the input from QDCO.

A major concern in a DDS based DTC implementation is the existence of spurious tones due to input-phase mismatches and non-linearities in the mixer input transistors. For instance, (\ref{eqn:uneq_cgd})  highlights that unequal gate-drain capacitance ($C_{GD}$) of mixer switches results in feedthrough of oscillator input (\textit{LO}) signal, thus generating spur of magnitude $V_x$ at fractional frequency ($f_{frac}$) offset from the desired DTC frequency.

\begin{equation}
\label{eqn:uneq_cgd}
V_{x} = V_{LO} \frac{C_{GD5}-C_{GD3}}{C_{GD1}+C_{GD6}+C_{mix^+}},
\end{equation}

\noindent where $C_{mix^+}$ is the total node capacitance at the drain of mixer switching transistor. Equation (\ref{eqn:frac2_eqn}) shows spur-generation at 2$f_{frac}$ offset from the DTC frequency. This spur occurs due to phase-mismatches ($\epsilon$) in the oscillator input (\textit{LO}), leading to incomplete rejection of the image-component  signal.

\begin{equation}
\label{eqn:frac2_eqn}
V_{out} = cos(\omega_{LO}-\omega_{frac})t + \epsilon\, cos(\omega_{LO}+\omega_{frac})t
\end{equation}

\begin{figure}[!h]
		\centering 
  		\includegraphics[trim={0cm 0.1cm 0cm 0cm},clip=true,scale=.27]{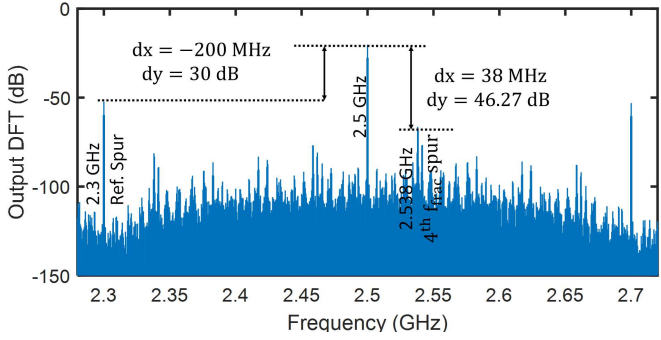}\\
  		 \caption{ Simulated frequency spectrum of the proposed dual-phase DDS based DTC for $f_{frac}$=9.5\,MHz. The mixer nonlinearities results in -47\,dBc spur at $4f_{frac}$ (=38\,MHz) offset.}
  		 \label{fig:DFT_Dual_DDS_based_DTC_with_MUX}
		\end{figure}

\begin{figure}[!h]
\centering
{
% \subfloat[]{
% \includegraphics[scale=0.3,trim={0.4cm 6cm 0cm 5cm}, clip]{././TF.pdf}}\\
% \subfloat[]{
% % \includegraphics[scale=0.48]{././INL_proposed.png}}
\includegraphics[scale=0.19]{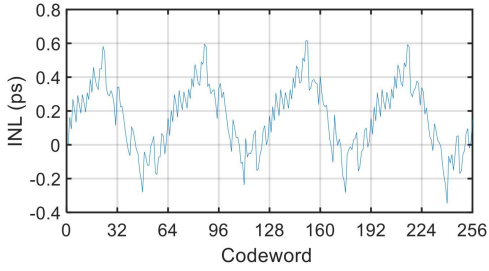}
}
\caption{Simulated INL of the dual-phase DDS based DTC, with $4f_{frac}$ spur governing the INL waveshape.}
\label{fig:dtc_tf_inl_simulated}
\end{figure}

Equation (\ref{eqn:uneq_cgd})-(\ref{eqn:frac2_eqn}) highlight that interconnect matching is crucial to avoid in-band spur-generation at the DTC output. Apart from the interconnect and device mismatches, the transconductance non-linearities of the mixer-switches results in generation of spurious tones at $4f_{frac}$ offset with respect to the output frequency. Figure \ref{fig:DFT_Dual_DDS_based_DTC_with_MUX} highlights the presence of spurious tone at $4f_{frac}$ offset from the desired signal. The spur-power level also governs the INL shape and magnitude, as observed from Fig. \ref{fig:dtc_tf_inl_simulated}.

%*************************************************************************************
%                            Switched-loop DPLL overview
%*************************************************************************************
\section{Switched-loop DPLL overview}
\label{ref:switched_loop}
\begin{figure}[!h]
\centerline{\includegraphics[scale = 0.4]{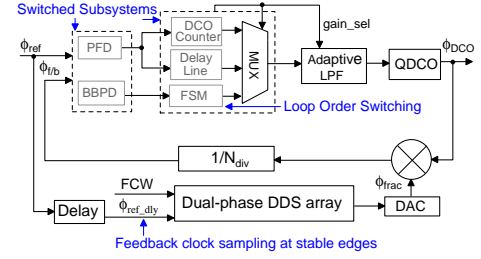}}
\caption{Detailed block diagram for the DPLL architecture.} 
\label{fig:pll_data_flow}
\end{figure}

\begin{figure}[!h]
\centerline{\includegraphics[scale = 0.58]{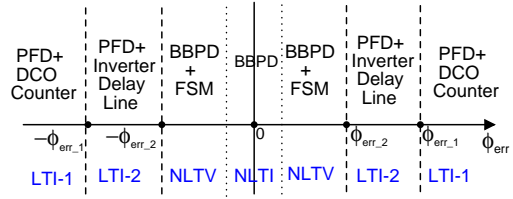}}
\caption{Phase-error state dependent switching rule for its subsystems.}
\label{fig:phase_err_fsm2}
\end{figure}

\begin{figure}[!h]
{
\subfloat[]{\includegraphics[scale = 0.35]{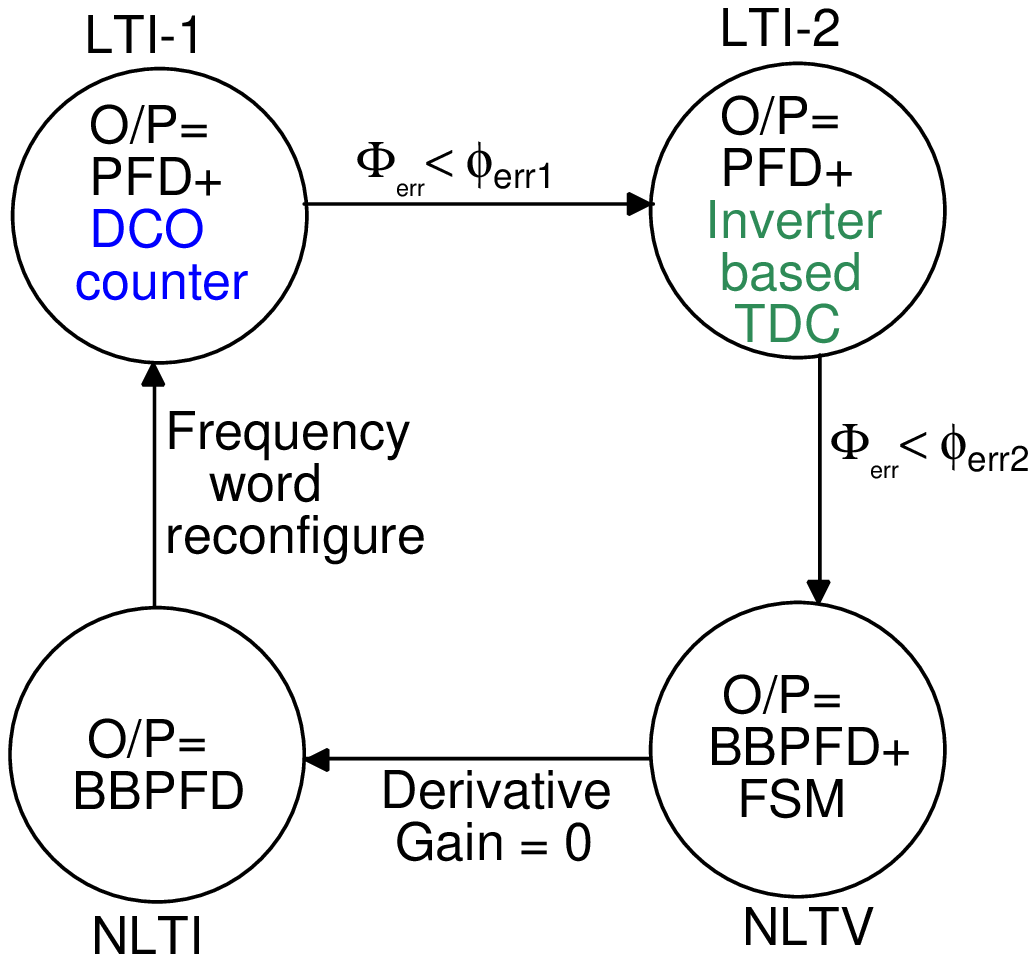}}
% \subfloat[]{\includegraphics[scale = 0.18]{./fsm41.eps}}
\subfloat[]{\includegraphics[scale = 0.35]{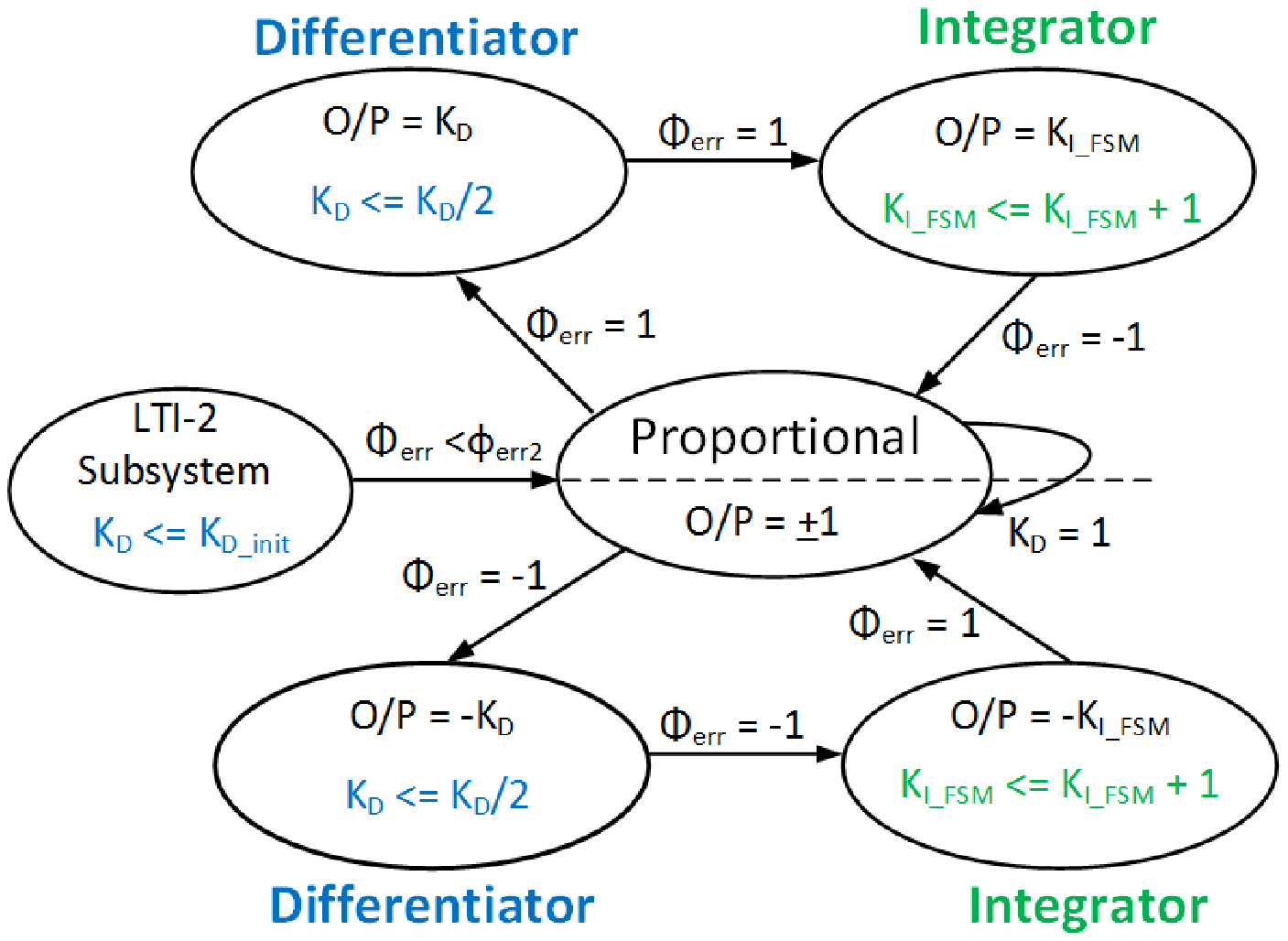}}
}
\caption{ (a) FSM for PFD+TDC variant activation based on the input phase error magnitude; (b) FSM algorithm activated with BBPD mode.}
\label{fig:fsm_diag}
\end{figure}

The DPLL \cite{ref:proposed_dpll} in Fig. \ref{fig:pll_data_flow} targets a low lock time-jitter product, by employing loop gain switching in the feedforward path and lookahead based phase interpolation in the feedback path.  This architecture involves switching between different subsystems based on the phase error state-dependent switching rule shown in Fig. \ref{fig:phase_err_fsm2}. Figure \ref{fig:fsm_diag}(a) shows that starting from a large phase error ($\phi_{err}$) magnitude, the loop traverses through activation of a linear phase frequency detector (PFD) with a DCO clock counter followed by switching to inverter based delay line. The deadzone in a single inverter ($\phi_{err2}$) is avoided by activating  bang-bang phase detection (BBPD). To improve the settling time, the BBPD is activated initially with a FSM emulating an additional  PID controller in the loop, as shown in Fig. \ref{fig:fsm_diag}(b). With the presence of a linear PFD, the system remains linear-time invariant (LTI); and becomes non-linear time variant (NLTV) or time invariant (NLTI) while switching to BBPD with or without FSM.

 When the hybrid phase detector enters bang-bang phase detection mode, the FSM gives a high initial derivative gain ($K_{D\_init}$) as correction on phase-error sign reversal. This derivative correction should be large enough to reduce the phase error below the value corresponding to an inverter delay ($\phi_{err\_2}$). If the BBPD asserts similar phase error sign in consecutive cycles, the FSM activates another integrator ($K_{I\_{FSM}}$) in the loop to achieve fast frequency tracking until the phase error sign changes. At every phase-error sign reversal, the FSM activates derivative gain for immediate phase alignment of the reference and feedback clock. The derivative gain ($K_D$) is reduced with each phase error sign reversal, assuming that the loop is undergoing settling process. When the derivative gain becomes 0, the FSM is removed from the loop to avoid chattering in the settled state. The fast-locking features in the loop's feedforward path stand ineffective in improving the settling response, if the system anyway has to spend time for calibrating the fractional divider in-between frequency switching. Towards this requirement, the DPLL employs the proposed DDS based DTC which inherits a calibration-free operation and an instantaneous frequency switching. 

%*************************************************************************************
% Measurement Results
%*************************************************************************************

\section{Measurement Results}
\label{sec:dtc_meas}

\begin{figure}[!h]
\centering{\includegraphics[scale=0.25]{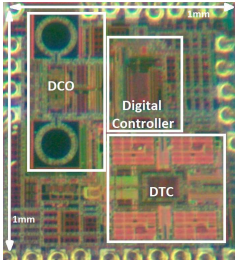}}
% \centering{\includegraphics[scale=0.25]{././micrograph_dtc_v3.eps}}
\caption{Chip micrograph of the DPLL incorporating the proposed DTC as a fractional divider.}
\label{fig:micrograph}
\end{figure}

\begin{figure}[!h]
\centering{\includegraphics[scale=0.21]{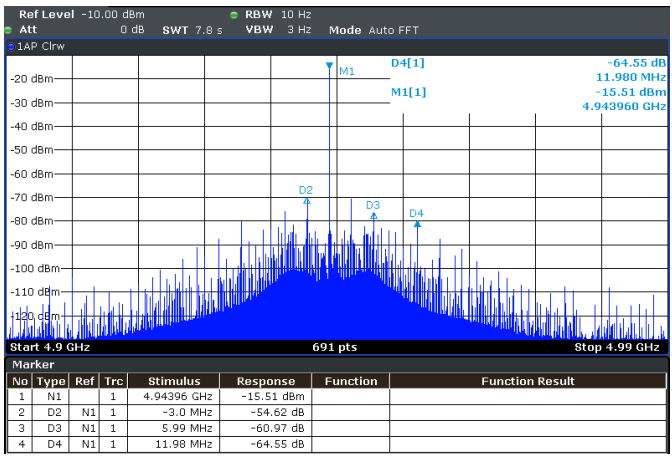}}
\caption{Measured output spectrum of DPLL with DDS programmed for $f_{frac}$=3\,MHz. The spurs generated at $f_{frac}$(=3\,MHz), $2f_{frac}$(=6\,MHz) and $4f_{frac}$(=12\,MHz) offsets are amplified by the loop bandwidth.} 
\label{fig:dpll_spectrum}
\end{figure}

The switched-loop DPLL incorporating the proposed DTC is implemented in CMOS\,65\,nm-LL technology with the chip micrograph as shown in  Fig. \ref{fig:micrograph}. The DPLL has a frequency range of 4.8-5\,GHz, while operating with 100\,MHz reference clock and 2\,MHz loop bandwidth in the settled state. The DTC operating directly at 5\,GHz DPLL output consumes 3\,mW power, with 2\,mW being consumed by current-weighted PI and 1\,mW being consumed by dual-phase DDS array. The measured spectrum of DPLL output in Fig. \ref{fig:dpll_spectrum} highlights the presence of spurious tones at $f_{frac}$, $2f_{frac}$ and $4f_{frac}$ offset as discussed in Section \ref{sec:dtc_implementation}. The magnitude of the spurs generated in DDS based DTC system is amplified by the loop bandwidth of the PLL. The DTC spur-level based on (\ref{eqn:in_band_spur}) impacts its INL response and thus the DPLL output jitter.

% \begin{figure}[!h]
% 
% \centering{
% \subfloat[]{\includegraphics[scale=0.13]{././chip_micrograph_2017_v2.eps}}
% \subfloat[]{\includegraphics[scale=0.3]{././dpll_jitter_histogram.eps}}}
% \caption{(a) Chip micrograph of the DPLL incorporating the proposed DTC as a fractional divider; (b) Measured jitter histogram of fractional-N DPLL with the jitter value governed by spur-level generated from the DTC.}
% \label{fig:micrograph}
% \end{figure}

\begin{figure}[!h]
\centering{\includegraphics[trim={0.4cm 4.5cm 0cm 4.5cm},clip=true,scale=.3]{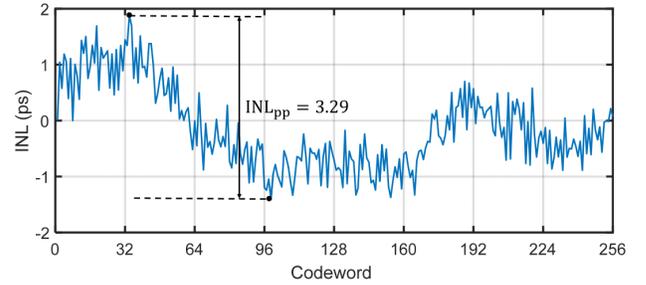}}
\caption{Measured DTC INL with the pattern governed by spur-level at $f_{frac}$, $2f_{frac}$ and $4f_{frac}$ frequency offsets.} 
\label{fig:dpll_inl}
\end{figure}

\begin{figure}[!h]
\centering{\includegraphics[scale=0.3]{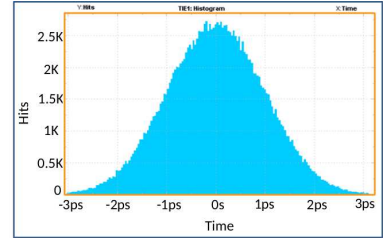}}
\caption{Measured jitter histogram of fractional-N DPLL with the jitter value governed by spur-level generated from the DTC.} 
\label{fig:dpll_jitter}
\end{figure}

\begin{figure}[!h]
\begin{center}
\includegraphics[scale=0.35]{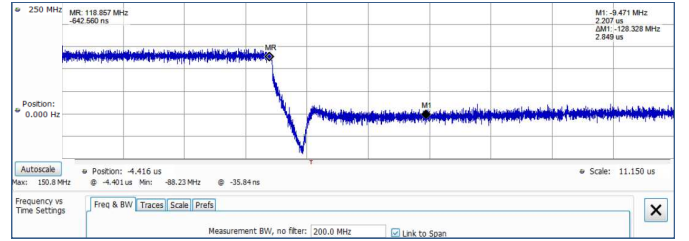}
\end{center}
\caption{Settling response  for fractional-N DPLL with 120\,MHz frequency step change.}
\label{fig:pll_settling_time}
\end{figure}

The proposed DDS based DTC with a phase-lookahead mechanism is able to achieve frequency-modulation in the range of $\pm$80\,MHz with the INL of 1.6\,ps, as shown in Fig. \ref{fig:dpll_inl}. Figure \ref{fig:dpll_jitter} presents jitter histogram of the DPLL output with the RMS jitter of 1\,ps. The DPLL jitter is in the range of 0.8\,ps-2\,ps, for the DDS output range of $\pm$80\,MHz, depending on whether the generated spurious tone is located in-band or out-of-band. In addition, being free from the calibration-loop convergence requirement, the DPLL with the proposed DTC achieves a fast lock time of 1\,$\mu$s, as shown in Fig. \ref{fig:pll_settling_time}.

%****************************************************************************
% Cancellation of spurious tone by predistorting digital frequency
%****************************************************************************
\begin{table*}[!t]
 \renewcommand{\arraystretch}{1.4}
    \caption{Performance comparison of various DTCs}
      \label{tab:dtc_cmp}
\begin{center}

\begin{tabular}{l |c c c c | c  c | c } 
\hline

           & \cite{ref:dtc_mux} & \cite{ref:dtc_hf }  & \cite{ref:dpll_dtc_pi} & \cite{ref:ILRO_phase_rotator} & \cite{ref:const_slope_dtc} & \cite{ref:ss_dtc_dpll}& \textbf{This Work} \\
          & ISSCC'16 & JSSC'13 & JSSC'16 & JSSC'17 & JSSC'15 & ESSCIRC'14  &  \\ \hline
Architecture & Contention-   &PI$+$Harmonic- & Pipelined & ILRO based &Constant-slope&DCDL with &   \textbf{Dual-Phase DDS}   \\
             &based PI&Rejection filter & PI& current-mode PI & DCDL &Switched Capacitor&\textbf{$+$current-mode PI}  \\
Technology (nm)     & 28      & 65   &  65 & 28           & 65   & 28    &  \bf{65}  \\ 
Frequency (GHz)     & 2.0     & 1.5  & 5   & 11           & 0.05 & 0.04 &  \bf{5.0}\\
Resolution [bit]    & 11      & 8    & 5   & 7            & 10   & 10& \bf{10}     \\
Resolution [Time] (ps)& 0.24  & 2.6  & -   & 0.7          & 0.18 & 0.55 & \bf{0.2} \\
Power (mW)         & 19.8     & 4.3  & 2.3 & 18          & 1.8  & 0.5 &  \bf{3}    \\
DNL (ps)           & 0.3      & 13   & -   & 0.35          &  -   & 0.44 &  \bf{0.39} \\
INL [pk] (ps)      & 1.2      & 19   & 1.5  & 0.77        & 0.15 & 0.45  &  \bf{0.25}  \\
Range    & Infinite& Infinite& Infinite& Infinite& 0.2\,ns & 0.56\,ns  &\bf{Infinite}\\
\hline
\end{tabular}
\end{center}
\end{table*}
\section{INL improvement with Calibration}
\label{sec:calibration}
To improve the INL at the DTC output, it is essential to cancel out the sinusoidal variation resulting from spur-genenration at $f_{frac}$,  $2f_{frac}$ and $4f_{frac}$ offsets. For this purpose, a foreground calibration based technique involving a strategic pre-distortion is applied to DDS-ROM with 10-bit phase-wordlength or address-lines. 

\begin{figure}[!h]
\centering
{
% \subfloat[]{
% % \includegraphics[trim={0.45cm 5cm 0cm 5cm},clip=true,scale=.32]{./INL_modeled.eps}
% }\\
% \subfloat[]{
\includegraphics[trim={0.4cm 5cm 0cm 5cm},clip=true,scale=.32]{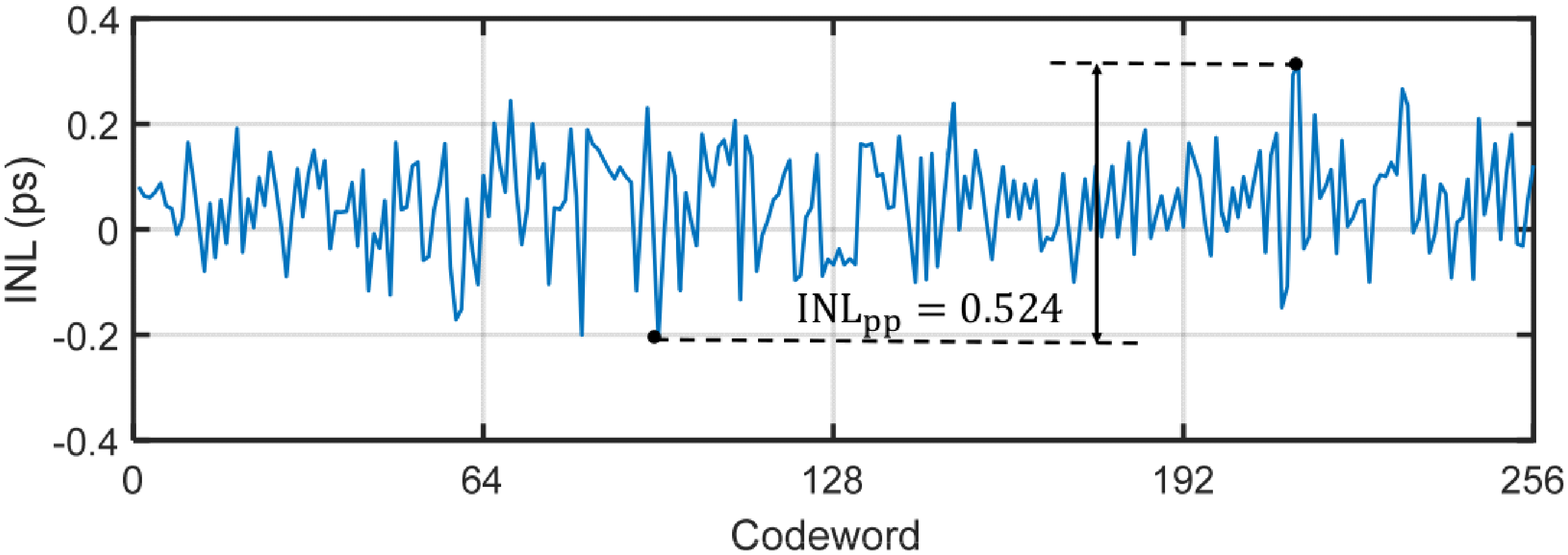}}
% }
\caption{Post-calibration improvement in the INL of proposed DTC.} 
\label{fig:post_calibration_inl}
\end{figure}

For verifying the effect of foreground calibration applied to the DDS-ROM,  the measured INL in Fig. \ref{fig:dpll_inl} is regenerated in simulation by modeling $C_{GD}$ mismatches in the mixer switches and phase-variations in the oscillator signal. The resultant INL after DDS-ROM pre-distortion in Fig. \ref{fig:post_calibration_inl} highlights the effectiveness of the foreground-calibration technique.  The peak INL of the DTC improves from 1.6\,ps  to 0.25\,ps, which also reflects as peak-to-peak jitter improvement from 1.2\,ps to 0.5\,ps at the  DTC output. (In the implemented fractional-N DPLL, only one stable edge out of N-edges at the DTC output is sampled by the system as the feedback clock.  Therefore, the DTC jitter is measured by observing the most stable edge out of N-edges being repetitively overlapped.)

% % % % % \begin{figure}[!h]
% % % % % \includegraphics[trim={0.1cm 0.05cm 0cm 0cm},clip=true,scale=.31]{./EyeDiagram-eps-converted-to.eps}
% % % % % \caption{Post-calibration jitter at the DTC output with the DDS programmed for 10\,MHz fractional frequency. } 
% % % % % \label{fig:post_calibration_jitter}
% % % % % \end{figure}

%****************************************************************************
% Performance Comparison for DTC
%****************************************************************************

\section{Performance Comparison}
\label{sec:perf_cmp}

Table I shows the performance comparison of different variants of DTC architecture. The calibrated dual-phase DDS based DTC system is able to achieve a competitive INL of 0.25\,ps with optimal power consumption, when compared to the GHz-domain phase interpolators allowing infinite delay-range over time. As illustrated in this work,  the proposed DDS$+$PI system finds its usage in fractional-N DPLLs which doesn't require all the edges of PI output to be stable, and samples only one edge out of N-edges of the DTC output. For such applications, conventional GHz-domain interpolators are an overkill with inclusion of power-consuming filtering blocks for reducing jitter at all the output edges.

 While the DCDL based DTCs achieve low INL with low power consumption, this performance is achieved with the limitations of (i) low operational frequency range and (ii) time-consuming background calibration slowing down the employing system's response. Thus for applications demanding frequency-translation with fast settling response, the proposed DDS$+$PI system turns out as a preferred solution over accumulator$+$DCDLs with limited range. For instance, the DPLL  incorporating the proposed fractional divider has a settling-time of 1\,$\mathrm{\mu}$s, while DCDL based DPLLs  need convergence-time of tens of microseconds for inital decision of coefficients used in the calibration technique.

\begin{figure}[!h]
\centering
	\includegraphics[scale = 0.5]{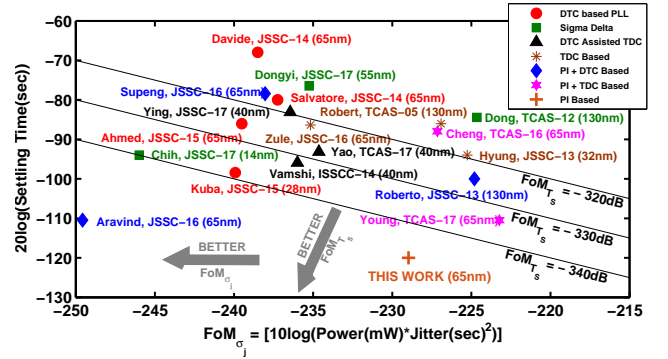}
\caption{Fractional-N DPLL performance benchmarking based on jitter, power and settling time response.}
\label{fig:new_fom}
\end{figure}

Figure \ref{fig:new_fom} highlights a competitive Figure of Merit (FoM) \cite{ref:dpll_fom} being achieved by the fractional-N DPLL incorporating the proposed phase-interpolator, in comparison to DPLLs employing calibration-dependent DTC/TDC as a fractional divider. The Figure of Merit used for DPLL performance benchmarking in Fig. \ref{fig:new_fom} is given as
\begin{equation}
\label{eqn:fom_adpll}
\mathrm{FoM} = 10\mathrm{log}\left[\left(\frac{\sigma_t}{1\mathrm{s}}\right)^2\left(\frac{t_s}{1\mathrm{s}}\right)^2\left(\frac{P}{1\mathrm{mW}}\right)\right].
\end{equation}

\section{Conclusion}

A dual-phase DDS based DTC system with phase-lookahead mechanism has been presented in this work. The proposed system achieves an extended frequency translation range, beyond the Nyquist rate of the DDS sampling clock. This DDS based DTC architecture, employed for fractional frequency shift in the feedback path of 5\,GHz DPLL, is implemented in CMOS65\,nm-LL technology with power consumption of 3\,mW. Without any background calibration requirement by this infinite delay-range DTC, the DPLL could achieve a fast settling response of 1\,$\mathrm{\mu}$s. Further linearization of the DTC has been shown with pre-distortion of the DDS-ROM based on the estimated nonlinearity. With this foreground calibration technique, the succeeding phase-interpolator achieves the best reported INL of 0.25\,ps, thus improving the jitter performance of the DPLL employing this system.

% The overall performance of the DPLL indicates that the system is able to achieve low settling time and jitter with the aid of the proposed DTC.
\bibliographystyle{IEEEtran}
\bibliography{ref}

\begin{thebibliography}{10}
\providecommand{\url}[1]{#1}
\csname url@rmstyle\endcsname
\providecommand{\newblock}{\relax}
\providecommand{\bibinfo}[2]{#2}
\providecommand\BIBentrySTDinterwordspacing{\spaceskip=0pt\relax}
\providecommand\BIBentryALTinterwordstretchfactor{4}
\providecommand\BIBentryALTinterwordspacing{\spaceskip=\fontdimen2\font plus
\BIBentryALTinterwordstretchfactor\fontdimen3\font minus
  \fontdimen4\font\relax}
\providecommand\BIBforeignlanguage[2]{{%
\expandafter\ifx\csname l@#1\endcsname\relax
\typeout{** WARNING: IEEEtran.bst: No hyphenation pattern has been}%
\typeout{** loaded for the language `#1'. Using the pattern for}%
\typeout{** the default language instead.}%
\else
\language=\csname l@#1\endcsname
\fi
#2}}

\bibitem{ref:dpll_dtc_pi}
A.~T. Narayanan, M.~Katsuragi, K.~Kimura, S.~Kondo, K.~K. Tokgoz, K.~Nakata,
  W.~Deng, K.~Okada, and A.~Matsuzawa, ``{A Fractional-N Sub-Sampling PLL using
  a Pipelined Phase-Interpolator With an FoM of -250\,dB},'' \emph{IEEE Journal
  of Solid-State Circuits}, vol.~51, no.~7, pp. 1630--1640, July 2016.

\bibitem{ref:dpll_pi}
Y.~H. Choi, B.~Kim, J.~Y. Sim, and H.~J. Park, ``A phase-interpolator-based
  fractional counter for all-digital fractional-n phase-locked loop,''
  \emph{IEEE Transactions on Circuits and Systems II: Express Briefs}, vol.~64,
  no.~3, pp. 249--253, March 2017.

\bibitem{Ref1_Extended_divider}
A.~Elkholy, S.~Saxena, R.~K. Nandwana, A.~Elshazly, and P.~K. Hanumolu, ``{A
  2.0--5.5 GHz Wide Bandwidth Ring-Based Digital Fractional-N PLL With Extended
  Range Multi-Modulus Divider},'' \emph{IEEE Journal of Solid-State Circuits},
  vol.~51, no.~8, pp. 1771--1784, 2016.

\bibitem{Ref7_vlsi_conf_dpll}
P.~Paliwal, J.~Fadadu, A.~Chawda, and S.~Gupta, ``{A Fast Settling 4.7-5 GHz
  Fractional-N Digital Phase Locked Loop},'' in \emph{IEEE International
  Conference on VLSI Design (VLSID)}, pp. 553--554, 2016.

\bibitem{Ref8_High_resolution_DTC}
A.~Chawda, P.~Paliwal, P.~Laad, and S.~Gupta, ``{High resolution
  digital-to-time converter for low jitter digital PLLs},'' in \emph{IEEE
  International Conference on Electronics, Circuits and Systems (ICECS)}, pp.
  25--28, 2014.

\bibitem{ref:proposed_dpll}
P.~Paliwal, J.~Fadadu, A.~Chawda, and S.~Gupta, ``{A Fast Settling 4.7-5 GHz
  Fractional-N Digital Phase Locked Loop},'' in \emph{International Conference
  on VLSI Design (VLSID)}, Jan 2016, pp. 553--554.

\bibitem{ref:dpll_fom}
P.~Paliwal, P.~Laad, M.~Sattineni, and S.~Gupta, ``Tradeoffs between settling
  time and jitter in phase locked loops,'' in \emph{IEEE International Midwest
  Symposium on Circuits and Systems (MWSCAS)}, Aug 2013, pp. 746--749.

\bibitem{Ref10_A_low_jitter-DTC}
H.~Sahu, P.~Paliwal, V.~Yadav, and S.~Gupta, ``{A low-jitter digital-to-time
  converter with look-ahead multi-phase DDS},'' in \emph{IEEE Latin American
  Symposium on Circuits \& Systems (LASCAS)}, pp. 219--222,2016.

\bibitem{ref:dds_pll}
A.~Bonfanti, D.~D. Caro, A.~D. Grasso, S.~Pennisi, C.~Samori, and A.~G.~M.
  Strollo, ``{A 2.5-GHz DDFS-PLL With 1.8-MHz Bandwidth in 0.35-$mu$m CMOS},''
  \emph{IEEE Journal of Solid-State Circuits}, vol.~43, no.~6, pp. 1403--1413,
  June 2008.

\bibitem{ref:dds_array}
T.~M. Comberiate, K.~C. Lauritzen, L.~B.~R. anf Cesar A.~Lugo, , and S.~H,
  ``{Spur Correlation in an Array of Direct Digital Synthesizers},''
  \emph{Proceedings of the 42nd Annual Precise Time and Time Interval Systems
  and Applications Meeting}, pp. 569 -- 584, November 2010.

\bibitem{ref:ulp_ss_dpll}
Y.~H. Liu, J.~V.~D. Heuvel, T.~Kuramochi, B.~Busze, P.~Mateman, V.~K. Chillara,
  B.~Wang, R.~B. Staszewski, and K.~Philips, ``{An Ultra-Low Power 1.7-2.7 GHz
  Fractional-N Sub-Sampling Digital Frequency Synthesizer and Modulator for IoT
  Applications in 40nm CMOS},'' \emph{IEEE Transactions on Circuits and Systems
  I: Regular Papers}, vol.~64, no.~5, pp. 1094--1105, May 2017.

\bibitem{ref:dtc_hf}
M.~S. Chen, A.~A. Hafez, and C.~K.~K. Yang, ``{A 0.1-1.5 GHz 8-bit
  Inverter-Based Digital-to-Phase Converter Using Harmonic Rejection},''
  \emph{IEEE Journal of Solid-State Circuits}, vol.~48, no.~11, pp. 2681--2692,
  Nov 2013.

\bibitem{ref:apf_pi}
K.~hyoun Kim, D.~M. Dreps, F.~D. Ferraiolo, P.~W. Coteus, S.~Kim, S.~V. Rylov,
  and D.~J. Friedman, ``{A 5.4mW 0.0035mm2 0.48psrms-jitter 0.8-to-5GHz
  non-PLL/DLL all-digital phase generator/rotator in 45nm SOI CMOS},'' in
  \emph{IEEE International Solid-State Circuits Conference - Digest of
  Technical Papers}, Feb 2009, pp. 98--99,99a.

\bibitem{ref:in_band_spur}
P.~Chen, X.~Huang, Y.~H. Liu, M.~Ding, C.~Zhou, A.~Ba, K.~Philips, H.~D. Groot,
  and R.~B. Staszewski, ``{Design and built-in characterization of
  digital-to-time converters for ultra-low power ADPLLs},'' in \emph{European
  Solid-State Circuits Conference (ESSCIRC)}, Sept 2015, pp. 283--286.

\bibitem{ref:octagonal_rotator}
G.~R. Gangasani, C.~M. Hsu, J.~F. Bulzacchelli, S.~Rylov, T.~Beukema,
  D.~Freitas, W.~Kelly, M.~Shannon, J.~Qi, H.~H. Xu, J.~Natonio, T.~Rasmus,
  J.~R. Guo, M.~Wielgos, J.~Garlett, M.~A. Sorna, and M.~Meghelli, ``{A 16-Gb/s
  Backplane Transceiver With 12-Tap Current Integrating DFE and Dynamic
  Adaptation of Voltage Offset and Timing Drifts in 45-nm SOI CMOS
  Technology},'' \emph{IEEE Journal of Solid-State Circuits}, vol.~47, no.~8,
  pp. 1828--1841, Aug 2012.

\bibitem{Ref11_time_interleaving}
S.~Balasubramanian, G.~Creech, J.~Wilson, S.~Yoder, J.~J. McCue, M.~Verhelst,
  and W.~Khalil, ``{Systematic analysis of interleaved digital-to-analog
  converters},'' \emph{IEEE Transactions on Circuits and Systems II: Express
  Briefs}, vol.~58, no.~12, pp. 882--886, 2011.

\bibitem{ref:dtc_mux}
S.~Sievert, O.~Degani, A.~Ben-Bassat, R.~Banin, A.~Ravi, B.~U. Klepser,
  Z.~Boos, and D.~Schmitt-Landsiedel, ``{A 2GHz 244fs-resolution
  1.2\,ps-Peak-INL edge-interpolator-based digital-to-time converter in 28\,nm
  CMOS},'' in \emph{IEEE International Solid-State Circuits Conference
  (ISSCC)}, pp. 52--54, 2016.

\bibitem{ref:ILRO_phase_rotator}
E.~Monaco, G.~Anzalone, G.~Albasini, S.~Erba, M.~Bassi, and A.~Mazzanti, ``{A
  2-11 GHz 7-Bit High-Linearity Phase Rotator Based on Wideband
  Injection-Locking Multi-Phase Generation for High-Speed Serial Links in 28nm
  CMOS FDSOI},'' \emph{IEEE Journal of Solid-State Circuits}, vol.~52, no.~7,
  pp. 1739--1752, July 2017.

\bibitem{ref:const_slope_dtc}
J.~Z. Ru, C.~Palattella, P.~Geraedts, E.~Klumperink, and B.~Nauta, ``{A
  High-Linearity Digital-to-Time Converter Technique: Constant-Slope
  Charging},'' \emph{IEEE Journal of Solid-State Circuits}, vol.~50, no.~6, pp.
  1412--1423, June 2015.

\bibitem{ref:ss_dtc_dpll}
N.~Markulic, K.~Raczkowski, P.~Wambacq, and J.~Craninckx, ``{A 10-bit, 550-fs
  step Digital-to-Time Converter in 28nm CMOS},'' in \emph{European Solid State
  Circuits Conference (ESSCIRC)}, Sept 2014, pp. 79--82.

\end{thebibliography}
\end{document}